\documentclass[aps,twocolumn,amsmath,amssymb,groupedaddress,floatfix]{revtex4-2}
\usepackage{graphicx}
\usepackage[english]{babel} 
\usepackage{bm}
\usepackage{xcolor}
\usepackage[normalem]{ulem}
\usepackage{booktabs}
\usepackage{hyperref}

\begin{document}

\title{Skimming transition in flexible granular sweeping}

\author{Yuto Ochi}
\author{Hiroaki Katsuragi}
\affiliation{Department of Earth and Space Science, The University of Osaka,
  1-1 Machikaneyama, Toyonaka 560-0043, Japan}

\date{\today}

\begin{abstract}
A flexible body placed in a steady flow bends to reduce drag. This self-streamlining is a hallmark of fluid-structure interaction (FSI). Granular-structure interaction is equally ubiquitous in nature. However, it remains poorly understood. Thus, we investigate inertial granular-structure interaction (IGSI). Specifically, ejection induced by a flexible plate sweeping a granular bed is experimentally examined. We find that a faster sweep results in less ejection, particularly for a flexible plate. To understand the underlying physics of this behavior, a dimensionless number Sk is introduced as the ratio of the plate elastic timescale to the sweep timescale. At $\mathrm{Sk} \lesssim 1$, the plate deflection follows the self-streamlining law of FSI and induces substantial ejection. At $\mathrm{Sk} \gtrsim 1$, on the other hand, the plate skims the bed and the mass of ejected grains decreases sharply. Sk organizes IGSI as the granular counterpart of FSI.
\end{abstract}

\maketitle

\section{Introduction}
Flexible structures sweep granular surfaces in many settings. For instance, an ostrich drags its wings to fling sand onto its body. Brooms, plows, and robot limbs exhibit similar behavior. These phenomena involve flexibility, inertia, transient contact, and free-surface ejection. We call this class of phenomena inertial granular-structure interaction (IGSI). In fluids, the corresponding class is fluid-structure interaction (FSI), and it is well developed. However, IGSI has no comparable framework, and no dimensionless quantity characterizing it has been proposed.

In fluid mechanics, the analogous problem has been studied within FSI~\cite{Langre:2008,Shelley:2011,Wang:2022}. A central result most relevant here is self-streamlining in steady flow. A flexible body deforms to align with the flow, and its drag falls below the rigid-body limit~\cite{Vogel:1984,Alben:2002,Alben:2004}. This mechanism explains how plants withstand wind and water~\cite{Vogel:1989,Gosselin:2010,Luhar:2011,Langre:2012,Gosselin:2019} and how insects achieve efficient flight with flexible wings~\cite{Miller:2009}. 

In granular physics, interactions with rigid bodies have been characterized over a wide velocity range. Granular drag in slow intrusion is described by resistive force theories~\cite{Li:2013,Askari:2016} or the frictional slip-line model~\cite{Kang:2018,Iikawa:2025}. At high velocities, granular media act as ejected momentum carriers, and impact-cratering scaling laws describe ejection mass and inertial drag~\cite{Housen:2011,Katsuragi:2016,Katsuragi:2007}. Flexible intruders have been considered only in the quasi-static regime~\cite{Algarra:2018,Pol:2025}, and jumping dynamics of deformable bodies on granular surfaces have also been studied~\cite{Aguilar:2016}. A systematic framework for flexible-structure sweeping of granular free surfaces is still missing. The key question is what timescale governs whether flexibility aids or hinders granular transport.

To solve this problem, we perform systematic experiments on granular sweeping by a flexible plate. Then, we introduce the skimming number Sk, the ratio of the elastic response timescale of the plate to the sweep timescale. We find that a threshold ($\mathrm{Sk} \simeq 1$) characterizes the whole behavior. Below the threshold, the plate deflection follows the self-streamlining scaling of FSI, although the sweep is transient and the medium is a granular bed with a free surface. Above the threshold, however, the plate skims the bed. As a consequence, the ejected mass falls as the sweep speeds up. Deformation, ejected mass, and ejection speed all turn over at the same threshold.

\section{Experiment}
\begin{figure}
\centering
\includegraphics[width=\linewidth]{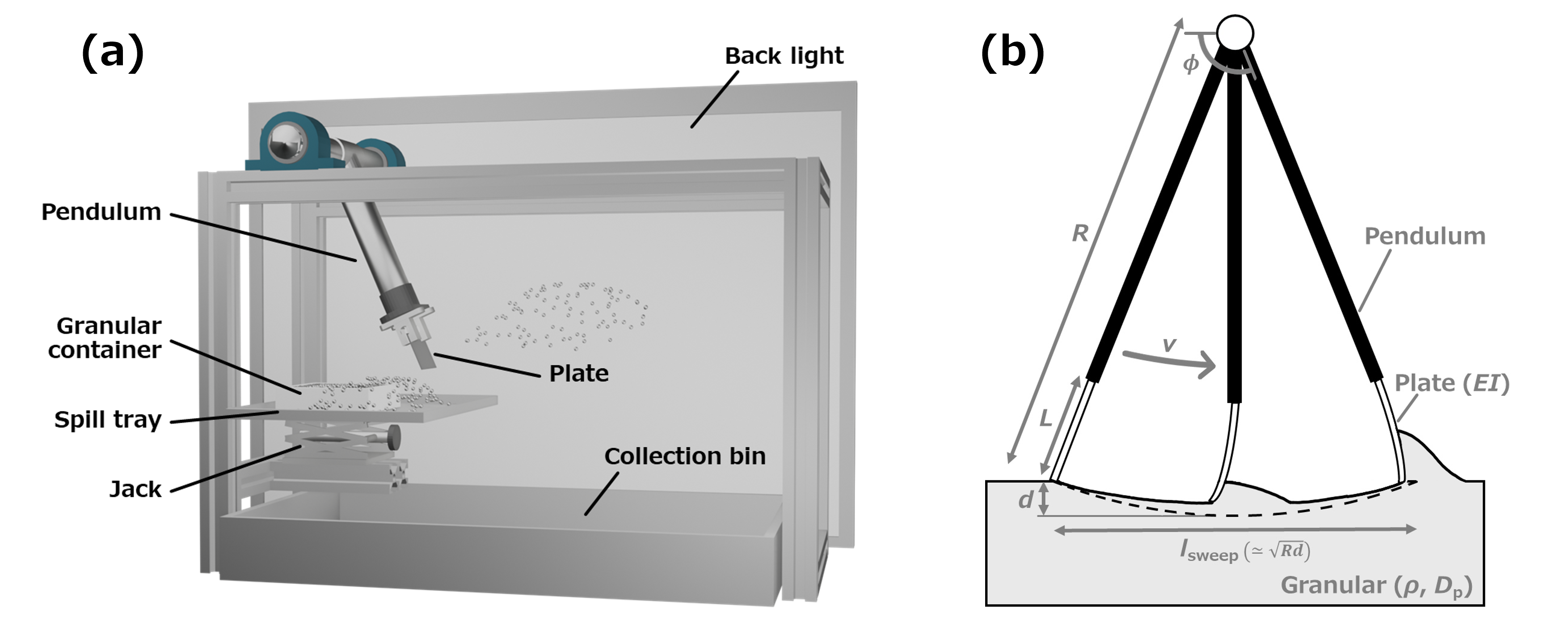}
\caption{Experimental setup. (a)~Three-dimensional schematic of the apparatus. A flexible plate attached to a pendulum sweeps a granular bed near the lowest point of the swing. (b)~Side-view schematic. Key parameters: pendulum radius $R$, plate length $L$, maximum penetration depth $d$, horizontal sweeping distance $l_{\mathrm{sweep}}$, and pendulum angle $\phi$ ($\phi = 0^\circ$ is horizontal; $\phi = 90^\circ$ is vertical).}
\label{fig:setup}
\end{figure}

The experimental apparatus consists of a pendulum arm ($R = 250$~mm) equipped with a flexible plate that sweeps a granular bed at controlled velocity $v$ and penetration depth $d$ (Fig.~\ref{fig:setup}). Three polycarbonate plates spanning two orders of magnitude in flexural rigidity were used: the Rigid plate ($EI = (8.41 \pm 0.10) \times 10^{-3}$~N\,m$^2$, thickness $1.02$~mm), the Flexible plate ($EI = (1.07 \pm 0.01) \times 10^{-3}$~N\,m$^2$, thickness $0.50$~mm), and the Highly flexible plate ($EI = (7.02 \pm 0.08) \times 10^{-5}$~N\,m$^2$, thickness $0.21$~mm). All plates share the same length $L = 30$~mm and width $w = 50$~mm. $EI$ was measured by static cantilever bending tests on a universal testing machine (Shimadzu AG-X) (Appendix~\ref{sec:EI}). Glass beads of four diameters ($D_\mathrm{p} = 2.0$, $1.0$, $0.8$, $0.2$~mm; bulk density $\rho = 1438 \pm 25$~kg\,m$^{-3}$) were used. Two penetration depths, $d = 1.0$ and $1.5$~mm, were tested. Forward-ejected mass $M$ was collected in a bin and measured with an electronic balance. Five independent runs were performed for each condition. 
The release angle of the pendulum $\phi_\mathrm{release}$ set the sweep velocity in the range $v = 1.5$--$3.3$~m\,s$^{-1}$. A high-speed camera (Photron FASTCAM Nova 20, 2000~fps, $0.36$~mm\,pixel$^{-1}$) recorded plate deformation and particle ejection. The pendulum arm was heavy enough ($0.87$~kg) that $v$ remained essentially constant during the sweep (Appendix~\ref{sec:v}). The instantaneous plate deformation $\delta(t)$ was measured and its maximum $\delta_{\mathrm{max}}$ during the sweep was extracted (Appendix~\ref{sec:delta}). The ejection speed $V$ was measured as the particle-cluster velocity immediately after its detachment from the plate (Appendix~\ref{sec:V}).

\section{Results}
\begin{figure}
\centering
\includegraphics[width=\linewidth]{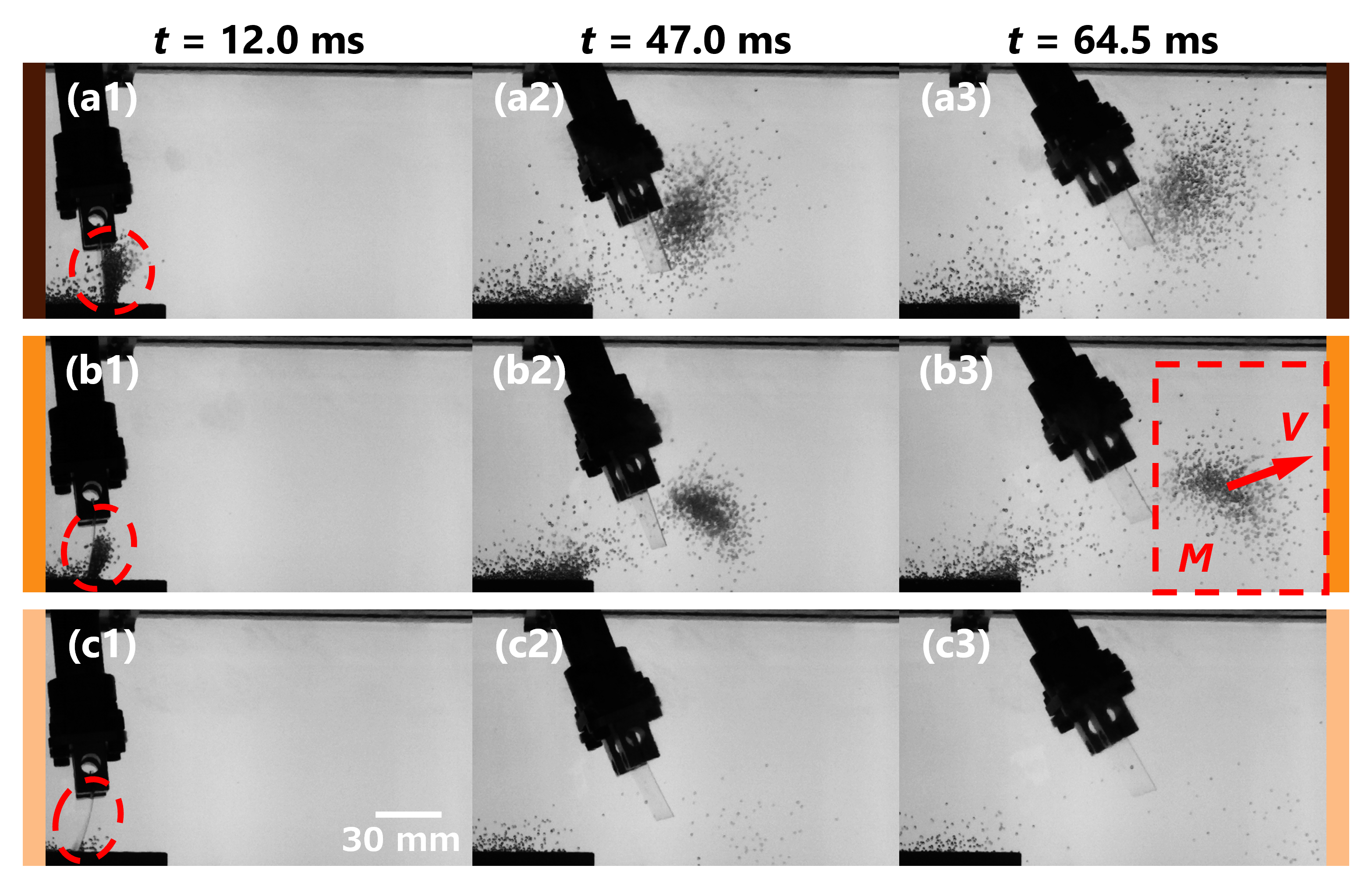}
\caption{Sequential snapshots at $t = 12.0$, $47.0$, and $64.5$~ms for (a1--a3)~the Rigid plate, (b1--b3)~the Flexible plate, and (c1--c3)~the Highly flexible plate, all at $v \simeq 2.0$~m\,s$^{-1}$. Orange dashed circles highlight plate deformation upon contact. The orange dashed rectangle in (b3) marks the region used to quantify $V$. Time origin $t = 0$ is at $\phi = 90^{\circ}$.}
\label{fig:rawdata}
\end{figure}

The three plates behave in visibly different ways (Fig.~\ref{fig:rawdata}). The Rigid plate captures and ejects particles diagonally upward. The Flexible plate deforms substantially. Particles accumulate near the deflected tip and are ejected as a compact cluster. The Highly flexible plate, in contrast, deforms extensively and glides over the surface. Forward ejection is suppressed almost entirely.

\begin{figure*}
\centering
\includegraphics[width=0.9\linewidth]{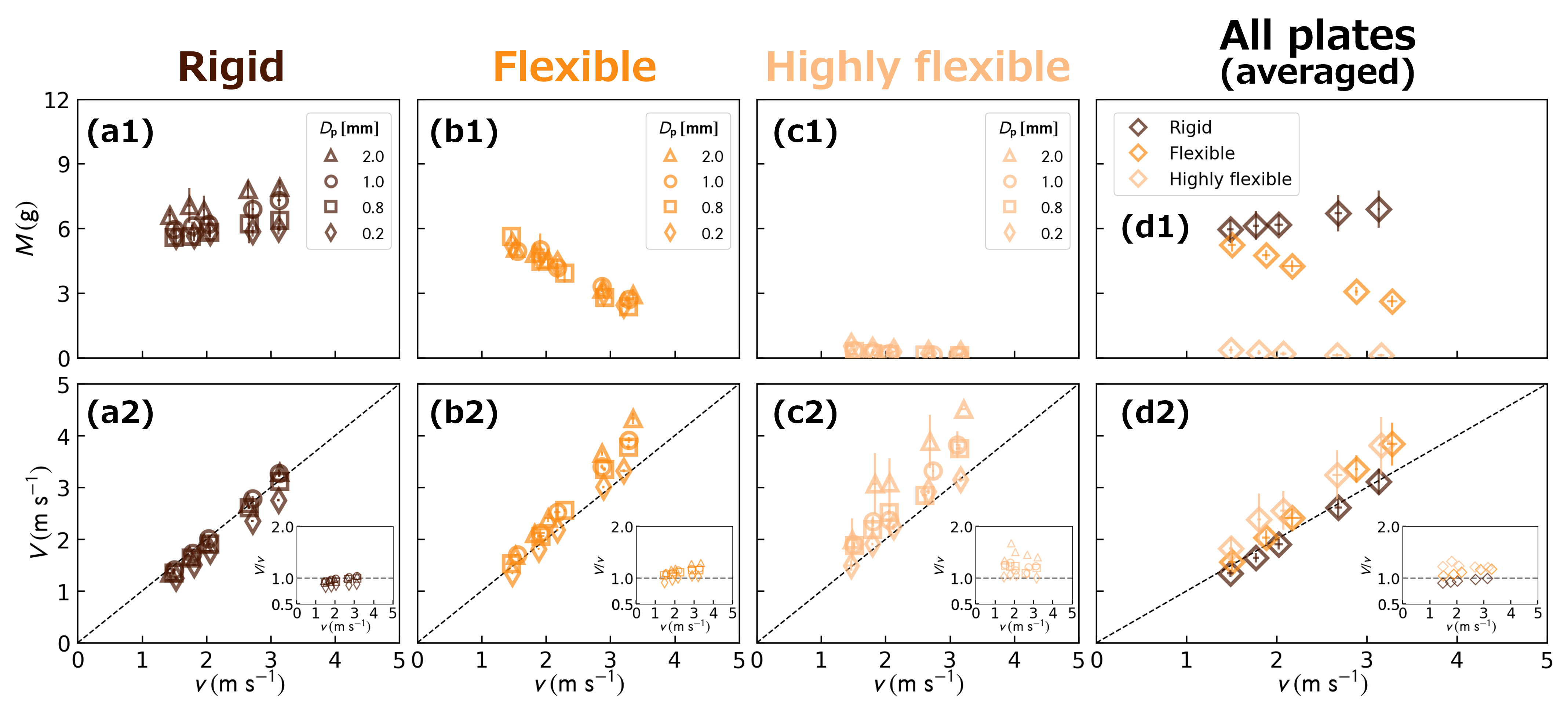}
\caption{Ejection parameters $M$ and $V$ versus $v$. Top row: $M$ versus $v$. Bottom row: $V$ versus $v$; insets show $V/v$. Columns (a)--(c): Rigid, Flexible, and Highly flexible plates. Markers distinguish $D_\mathrm{p} = 2.0$, $1.0$, $0.8$, $0.2$~mm. Column (d): $D_\mathrm{p}$-averaged data on common axes. Panel (d1) highlights the reversal of the $M$-$v$ trend. Panel (d2) and inset show $V/v > 1$ for flexible plates. Dashed lines mark $V = v$. All data: $d = 1.0$~mm. Error bars: standard deviation of $5$ trials.}
\label{fig:MV}
\end{figure*}

Figure~\ref{fig:MV} shows measured $M$ and $V$ versus $v$ for each plate. For the Rigid plate, $M$ increases monotonically with $v$. For the Flexible plate, however, $M$ decreases monotonically with $v$. For the Highly flexible plate, $M$ is very small throughout the velocity range. This reversal is the central puzzle of this study. A simple momentum argument predicts only that a softer plate transfers less momentum at fixed $v$. It does not predict that the transported mass falls as the sweep speeds up.

The ejection speed $V$ reveals another aspect of IGSI (Fig.~\ref{fig:MV}(a2)--(d2)). For the Rigid plate, $V/v \simeq 1$, consistent with the rigid-body limit. For the Flexible and Highly flexible plates, $V/v$ exceeds unity at high $v$. The trends of both $M$ and $V$ are insensitive to particle size $D_\mathrm{p}$. Despite a tenfold variation in $d/D_\mathrm{p}$, data for different $D_\mathrm{p}$ collapse onto one another, validating a continuum-level description of IGSI. The plate stores elastic energy during the sweep and converts it into particle kinetic energy upon detachment. This is the \emph{slingshot effect}. The flexible plate thus moves less material while throwing it faster. A single mechanism must produce both trends.

\section{Analysis}
The reversal of the $M$-$v$ relationship implies a competition between two timescales. The first is the elastic response time $t_{\mathrm{plate}}$ of the plate loaded by the granular bed. The measured $\delta(t)$ is a single damped oscillation (Appendix~\ref{sec:delta}), so we model the plate as a cantilever with stiffness $k = 3EI/L^3$. The effective tip mass is the granular material mobilized ahead of the plate, $M_{\mathrm{add}} \sim \rho w^2 d$, where the forward reach of that region is set by the plate width (Appendix~\ref{sec:Madd}). The plate mass is much smaller than $M_{\mathrm{add}}$ for all three plates, so the plate inertia is dropped to leading order. Thus, 
$t_{\mathrm{plate}} \sim \sqrt{M_{\mathrm{add}}/k} \sim \sqrt{\rho w^2 d L^3/(EI)}.$ 
The second is the sweep time $t_{\mathrm{sweep}}$. For a pendulum of radius $R$ penetrating to depth $d$, the contact distance is $l_{\mathrm{sweep}} \simeq \sqrt{Rd}$, so 
$t_{\mathrm{sweep}} \sim \sqrt{Rd}/v.$
Then we define the skimming number as the ratio 
\begin{equation}
\mathrm{Sk} = \frac{t_{\mathrm{plate}}}{t_{\mathrm{sweep}}} \sim \sqrt{\frac{\rho v^2 w^2 L^3}{EI R}}.
\label{eq:Sk}
\end{equation}
The penetration depth $d$ cancels in the shallow-sweeping regime ($d \ll L$ and $d \ll R$). This cancellation is a prediction rather than an assumption. Written in the general form $M_{\mathrm{add}} \sim \rho w \ell d$, with $\ell$ the forward extent of the mobilized region, the same steps give $\mathrm{Sk} \sim \sqrt{\rho v^2 w \ell L^3 / (EI R)}$. The depth drops out only when $\ell$ is set by the plate and not by the sweep. If $\ell \sim \sqrt{Rd}$ or $\ell \sim d$, $d$ would survive and data taken at different depths would not fall on one curve. 

Once the $d$ dependence is removed, Sk becomes proportional to $\mathrm{Ca}^{1/2}$, where Ca is the Cauchy number of FSI. Sk is not, however, a repackaged Ca. Ca is a ratio of forces and applies to steady states. Sk is a ratio of timescales and applies to transient contact. Since IGSI is transient, the timescale ratio is the natural governing quantity.

\begin{figure}
\centering
\includegraphics[width=0.9\linewidth]{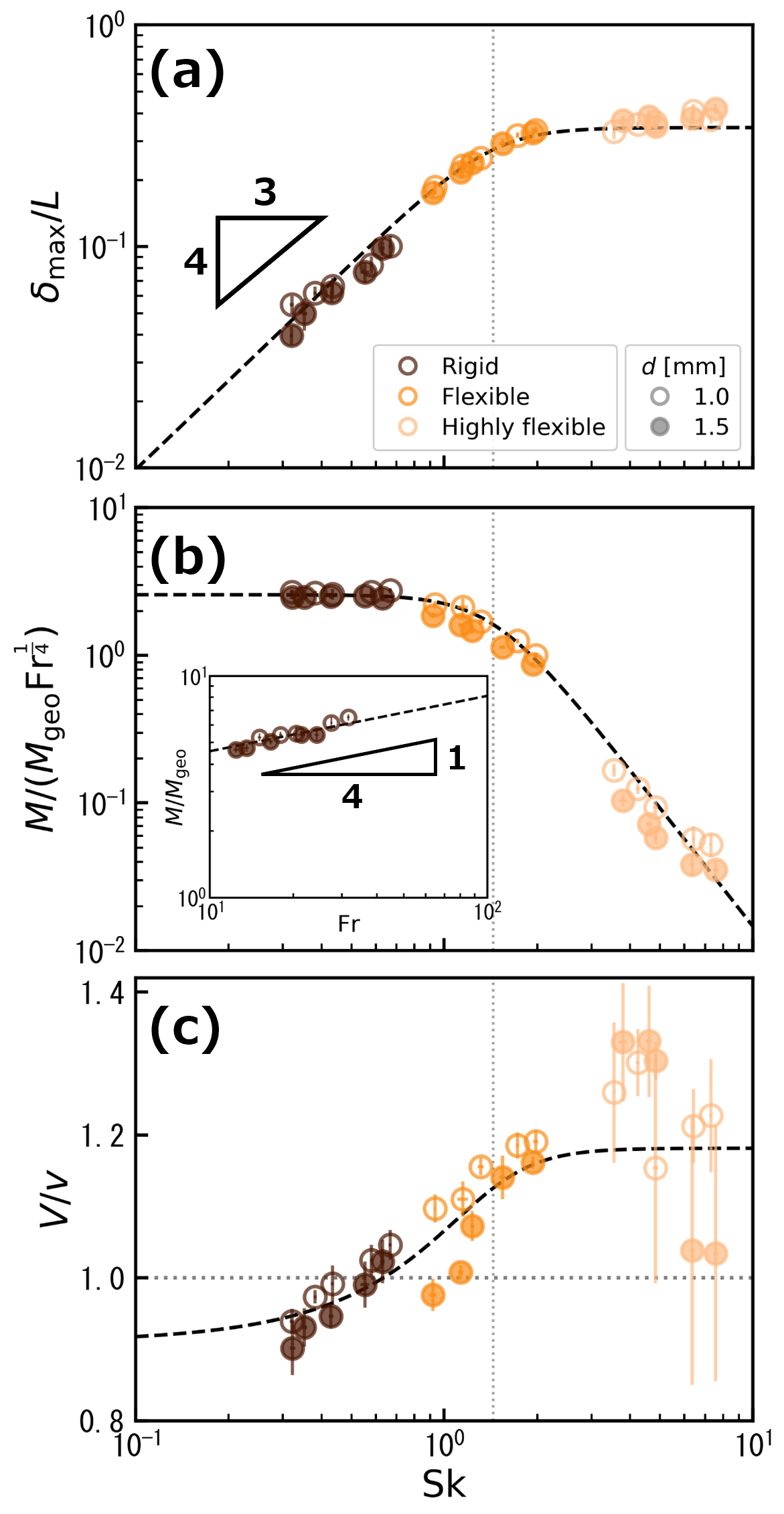}
  \caption{Scaling of three observables against Sk, with a common crossover scale $\mathrm{Sk_c} = 1.4$. (a)~Normalized maximum deformation $\delta_{\mathrm{max}}/L$. Dashed curve: Eq.~\eqref{eq:delta} with $C_\delta = 0.34$. (b)~Normalized ejected mass $M/(M_{\mathrm{geo}} \mathrm{Fr}^{1/4})$. Dashed curve: Eq.~\eqref{eq:M} with $C_M = 2.6$. Inset: Rigid-plate baseline $M/M_{\mathrm{geo}}$ against $\mathrm{Fr} = v/\sqrt{gd}$. Dashed line is $\mathrm{Fr}^{1/4}$. (c)~Velocity ratio $V^*=V/v$. Dashed curve: Eq.~\eqref{eq:V} with $V^*_0 = 0.91$ and $\Delta V^* = 0.27$. Colors match Fig.~\ref{fig:MV}. Open and filled symbols denote $d = 1.0$ and $1.5$~mm, respectively. Data shown: $D_\mathrm{p} = 1.0$~mm. Error bars: standard deviation of $5$ trials.}
\label{fig:scaling}
\end{figure}

Figure~\ref{fig:scaling} shows all three observables plotted against Sk. Each collapses to a single curve over two orders of magnitude in $EI$ and over both penetration depths. The collapse of data taken at different values of $d$ is consistent with the $d$ independence of Sk discussed above (Eq.~\eqref{eq:Sk}). 

For $\mathrm{Sk} \lesssim 1$, $\delta_{\mathrm{max}}/L \sim \mathrm{Sk}^{4/3}$ (Fig.~\ref{fig:scaling}(a)). This exponent is identical to the hallmark of FSI self-streamlining. It follows from dimensional analysis. Let $f$ be the drag force per unit width, $\rho$ the medium density, $v$ the velocity, and $B = EI/w$ the bending stiffness per unit width. These quantities combine in only one way,
\begin{equation}
f \propto B^{1/3} \left( \rho v^2 \right)^{2/3},
\label{eq:43_force}
\end{equation}
identical to the FSI result~\cite{Gosselin:2010}. Inserting into $\delta \sim f L^3 / B$ yields $\delta/L \sim \mathrm{Sk}^{4/3}$ directly. This argument requires only that the medium exert an inertial stress set by a density and a velocity. It does not require a steady flow, a continuous medium, or the absence of a free surface. The same exponent is therefore expected whenever the granular bed acts as an effective continuum. The observed insensitivity to $D_\mathrm{p}$ and $d$ supports that description. At $\mathrm{Sk} \gtrsim 1$, however, the deformation saturates. The bent plate tip can no longer scoop the grains. It merely skims the surface. A single crossover function connects the two limits,
\begin{equation}
  \frac{\delta_{\mathrm{max}}}{L} = C_\delta \, \frac{(\mathrm{Sk}/\mathrm{Sk_c})^{4/3}}{\left[1 + (\mathrm{Sk}/\mathrm{Sk_c})^{4}\right]^{1/3}}.
\label{eq:delta}
\end{equation}
Here, $C_{\delta}=0.34$ is the saturated value of $\delta_{\mathrm{max}}/L$ at $\mathrm{Sk} \gg \mathrm{Sk_c}$, and $\mathrm{Sk_c}=1.4$ is the crossover scale. The dashed curve in Fig.~\ref{fig:scaling}(a) displays Eq.~\eqref{eq:delta}. We call the crossover near $\mathrm{Sk} \simeq 1$ the \emph{skimming transition}.

Two factors set the ejected mass $M$. One is a velocity-dependent baseline that survives even for the Rigid plate. The other is the suppression above $\mathrm{Sk_c}$. We normalize $M$ by the geometric reference mass $M_{\mathrm{geo}}$, the mass contained in the circular segment swept by the plate at depth $d$ (Appendix~\ref{sec:Mgeo}). Whereas $M_{\mathrm{geo}}$ itself is $v$-independent, the Rigid plate shows $M$ increasing with $v$ (Fig.~\ref{fig:MV}(a1)). We attribute this residual velocity dependence to the Froude number $\mathrm{Fr} = v/\sqrt{gd}$ ($g=9.8$~m\,s$^{-2}$: gravitational acceleration), since the competition between inertia and gravity governs this limit. As shown in the inset of Fig.~\ref{fig:scaling}(b), $M/M_{\mathrm{geo}} \sim \mathrm{Fr}^{1/4}$ for the Rigid plate. 
We now turn to the second factor. Above $\mathrm{Sk_c}$, $M/M_\mathrm{geo}$ is suppressed because the plate fails to keep satisfying the FSI-like deformation law $\delta_\mathrm{max}/L \sim (\mathrm{Sk/Sk_c})^{4/3}$. Thus, this deformation-reduction factor is proportional to $\left[1 + (\mathrm{Sk}/\mathrm{Sk_c})^{4}\right]^{-1/3}$. The plate responds linearly. So this same factor also gives the ratio between the actual load on the grains and the load required for unabated $\mathrm{Sk}^{4/3}$ growth. The mass set into motion, however, is governed by the work done on the grains rather than by the load alone. For a linear response, this work scales as the square of the load ratio. Squaring the reduction factor gives
\begin{align}
  M &= C_M M_{\mathrm{geo}} \mathrm{Fr}^{1/4} \left[1+(\mathrm{Sk}/\mathrm{Sk_c})^{4}\right]^{-2/3} \nonumber \\
    &\sim M_{\mathrm{geo}} \mathrm{Fr}^{1/4} \left( \frac{\delta_\mathrm{max}}{L} \right)^2 \mathrm{Sk}^{-8/3},
\label{eq:M}
\end{align}
with $C_M=2.6$. This function describes the data well~(dashed curve in Fig.~\ref{fig:scaling}(b)).
We find $C_M > 1$. This is because $M_{\mathrm{geo}}$ counts only the material inside the swept segment while the plate also drives more grains.

At high Sk, fewer particles leave the plate but at higher speed (Fig.~\ref{fig:scaling}(c)). Since $\delta$ is bounded, $V^* = V/v$ must saturate at large Sk. We consider that the slingshot boost is caused by the momentum delivered to the departing grains at the instant of detachment. Then, it is proportional to $\delta_{\mathrm{max}}/L$. With the same crossover shape as Eq.~\eqref{eq:delta} for $\delta_{\mathrm{max}}/L$, the ejection speed follows
\begin{equation}
  V^* - V^*_0 = \Delta V^* \frac{(\mathrm{Sk}/\mathrm{Sk_c})^{4/3}}{\left[1 + (\mathrm{Sk}/\mathrm{Sk_c})^{4}\right]^{1/3}} \sim \frac{\delta_\mathrm{max}}{L},
\label{eq:V}
\end{equation}
with two fitting parameters $V^*_0=0.91$, $\Delta V^*=0.27$~(dashed curve in Fig.~\ref{fig:scaling}(c)). At $\mathrm{Sk} \ll 1$, $V^* \to V^*_0$. The fitted value $V^*_0$ falls below the ideal rigid-body value of unity. This reduction reflects dissipation in the granular medium. All three observables, $\delta_\mathrm{max}$, $M$, and $V$, can be rationalized by the identical crossover function.

\section{Discussion}
The skimming transition at $\mathrm{Sk} \simeq 1$ appears as the reversal of the $M$-$v$ trend for the Flexible plate. At this point, elastic deformation outpaces the sweep and the plate ceases to scoop. A single quantity Sk captures both the FSI-like scaling regime and the granular-specific skimming transition. For $\mathrm{Sk} \lesssim 1$, the system follows the $4/3$ scaling shared with continuum FSI~\cite{Gosselin:2010}. This correspondence suggests that self-streamlining may extend beyond conventional fluids into granular environments.

For $\mathrm{Sk} \gtrsim 1$, however, IGSI departs from the classical FSI picture. In this regime, the transported mass turns over and falls with sweep speed. The ejection speed rises above the sweep speed. Both changes take place around a single threshold, $\mathrm{Sk}\simeq 1$.

This transition has practical implications. Devices that interact with granular surfaces should be designed to keep $\mathrm{Sk} \lesssim 1$ if transport efficiency is required. Potential applications include sampling tools on planetary rovers, robot limbs for desert travel, and agricultural tillage equipment. Conversely, for tasks where ground disturbance must be minimized, such as sampling fragile sediments, the skimming regime at $\mathrm{Sk} \gg 1$ can be used intentionally.

Beyond engineering applications, the same perspective applies to biology. Ostrich sand-bathing, in which a flightless bird actively flings granular material with its wings, is a natural example of IGSI. The present framework allows us to analyze physical behaviors using measurable quantities, such as wing rigidity, motion speed, and particle size. By computing Sk from these quantities, we can determine which regime these behaviors fall into. Comparing wing morphologies among related taxa may then reveal how evolution has optimized granular dispersal.

The current IGSI framework has several limitations. First, the experiment uses a quasi-two-dimensional geometry. Second, the experiment and model can apply only to shallow sweeping with constant $v$. Third, $w$, $L$, and $R$ were fixed, so their exponents in Eq.~\eqref{eq:Sk} mainly rely on dimensional analysis. These issues are left for future work.

Even within these bounds, a single timescale ratio organizes the reversal of mass and speed trends as it crosses unity. It places granular sweeping and fluid-structure interaction within a common framework. Whether the same elastic-inertial competition governs other transient interactions with yielding media remains an open question we find worth pursuing.

\begin{acknowledgments}
This work was partially supported by JSPS KAKENHI Grant No.~JP24H00196 and JST ERATO Grant No.~JPMJER2401. 
The data that support the findings of this study are openly available in Zenodo at Ref.~\cite{Ochi_data:2026}. 
During the preparation of this work, the authors used Gemini, ChatGPT, and Claude in order to assist writing, coding, and discussion. After using these tools, the authors reviewed and edited the content as needed and take full responsibility for the content of the publication.
\end{acknowledgments}

\appendix
\setcounter{secnumdepth}{4}

\section{Flexural rigidity \texorpdfstring{$EI$}{EI} measurement}
\label{sec:EI}
\textbf{Experimental conditions.} The flexural rigidity $EI$ of each plate was evaluated through static cantilever bending tests using a universal testing machine (AG-X, Shimadzu Corp.). To ensure reproducibility, three independent measurements were performed for each specimen, with each trial including the complete mounting and dismounting of the plate from the fixture. A line load covering the full plate width ($w = 50$~mm) was applied at a distance $L_{\mathrm{b}} = 27$~mm from the fixed end. The reaction force $F_{\mathrm{b}}$ and displacement $d_{\mathrm{b}}$ were recorded at a constant indentation speed of $1.0$~mm~min$^{-1}$.

\begin{figure}[htbp]
    \centering
    \includegraphics[width=1\linewidth]{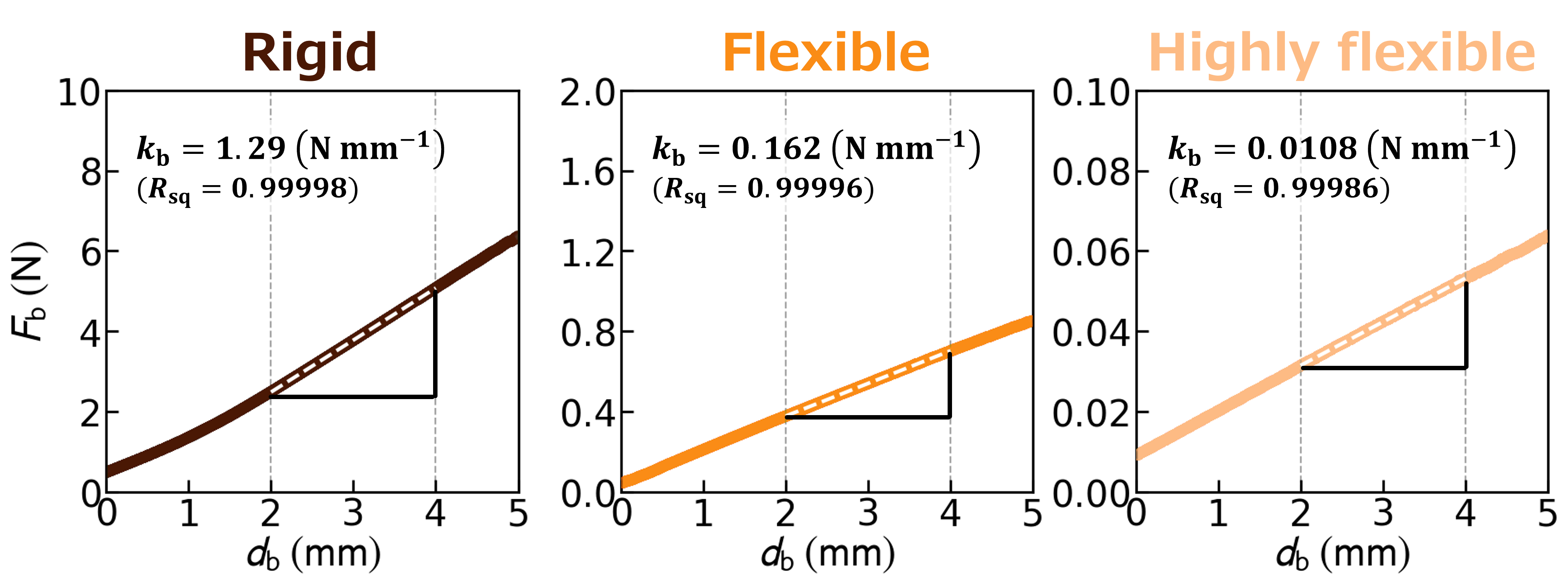}
    \caption{Reaction force $F_{\mathrm{b}}$ versus displacement $d_{\mathrm{b}}$ for each plate. The displacement range from $2.0$ to $4.0$~mm (indicated between the vertical dotted lines) was treated as the linear region to calculate the stiffness coefficient $k_{\mathrm{b}}$ via the least-squares method. The values of $R_{\mathrm{sq}}$ in parentheses indicate the coefficients of determination. Note that the values of $k_{\mathrm{b}}$ shown in the plots are obtained from a single representative trial, whereas Table~\ref{tab:EI_results} reports the ensemble averages over three independent trials.}
    \label{fig:EI}
\end{figure}

\textbf{Data analysis.} Fig.~\ref{fig:EI} shows representative load-displacement curves. To eliminate initial contact misalignment and non-linear settling behaviors, the displacement range from $2.0$ to $4.0$~mm was defined as the linear region. Within this interval, the stiffness coefficient $k_{\mathrm{b}} = \Delta F_{\mathrm{b}} / \Delta d_{\mathrm{b}}$ was calculated using the least-squares method. The flexural rigidity $EI$ was then determined based on linear beam theory,
\[EI = \frac{k_{\mathrm{b}} L_{\mathrm{b}}^3}{3}.\]
The validity of the calculated $EI$ value was verified by converting it to Young's modulus $E$ using the second moment of area $I = wh^3/12$,
\[E = \frac{EI}{I}.\]
The obtained values were then compared with the literature value to confirm consistency.

\textbf{Results.} Table~\ref{tab:EI_results} summarizes the measured mechanical properties and bending stiffness for the plates of each thickness. The effective thickness $h$ was precisely measured using a micrometer (MDC-25M, Mitutoyo Corp.). The estimated $E$ agreed with the literature value to within $18\%$ for all specimens ($= 2.2$~GPa; Technical Data for Plastic Washers/Collars, MISUMI Group Inc., 2021). Despite this slight discrepancy, the calculated Young's modulus exhibited no significant thickness dependence or systematic variation across the plates. Consequently, the bending stiffness derived from these direct measurements was adopted for physical analyses in this study.

\begin{table*}[htbp]
\centering
\caption{Measured bending stiffness and derived mechanical properties of the polycarbonate plates. The tests were conducted with a common width $w = 50$~mm and an effective length $L_{\mathrm{b}} = 27$~mm.}
\begin{tabular}{lcccc}
\toprule
 & $h$ ($\mathrm{mm}$) & $k_{\mathrm{b}}$ ($\mathrm{N~m^{-1}}$) & $EI$ ($\mathrm{N~m^2}$) & $E$ ($\mathrm{GPa}$)\\
\midrule
Rigid           & 1.02 & $(1.28 \pm 0.01) \times 10^{3}$ & $(8.41 \pm 0.10) \times 10^{-3}$ & $1.90 \pm 0.02$ \\
Flexible        & 0.50 & $(1.62 \pm 0.02) \times 10^{2}$ & $(1.07 \pm 0.01) \times 10^{-3}$ & $2.05 \pm 0.03$ \\
Highly flexible & 0.21 & $(1.07 \pm 0.01) \times 10^{1}$ & $(7.02 \pm 0.08) \times 10^{-5}$ & $1.82 \pm 0.02$ \\
\bottomrule
\end{tabular}
\label{tab:EI_results}
\end{table*}

\section{Sweeping velocity \texorpdfstring{$v$}{v} measurement}
\label{sec:v}
\begin{figure}[htbp]
    \centering
    \includegraphics[width=1\linewidth]{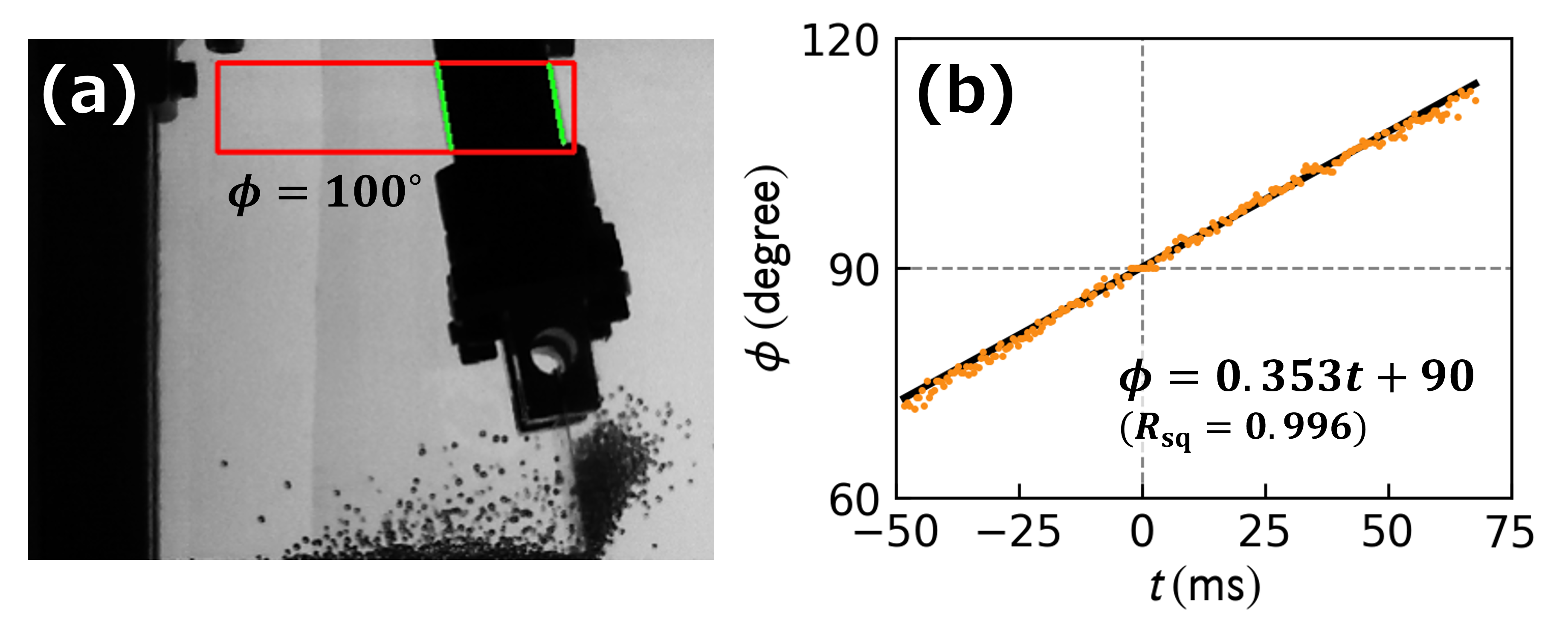}
    \caption{Measurement of the pendulum angle $\phi$. (a)~Image analysis for determining the pendulum angle $\phi$. Green lines indicate the detected edges of the pendulum arm within the red region of interest. When both front and back edges are detected, $\phi$ is calculated as their average orientation. (b)~Temporal evolution of the pendulum angle $\phi$ for $D_{\mathrm{p}} = 1.0$~mm and $v \simeq 1.55$~m~s$^{-1}$. The experimental data show high linearity ($R_{\mathrm{sq}} = 0.996$) around the granular impact point ($t = 0$, $\phi = 90^\circ$, indicated by dashed lines), validating the assumption of constant sweeping velocity during the interaction.}
    \label{fig:phi}
\end{figure}

The angular displacement of the pendulum $\phi(t)$ was extracted from high-speed video frames ($2000$~fps) using a line-detection algorithm based on the Hough transform. As displayed in Fig.~\ref{fig:phi}, the angular displacement $\phi$ exhibits excellent linearity, even during the high-resistance regime where the plate sweeps through the granular bed ($t \simeq 0$). The high coefficient of determination ($R_{\mathrm{sq}} = 0.996$) demonstrates that the deceleration induced by the granular medium is negligible relative to the large rotational inertia of the pendulum system. Therefore, the tangential sweeping velocity $v$ at the plate tip was treated as a constant throughout the entire interaction process.

\section{Plate deformation \texorpdfstring{$\delta$}{delta} measurement}
\label{sec:delta}
\begin{figure}[htbp]
    \centering
    \includegraphics[width=1\linewidth]{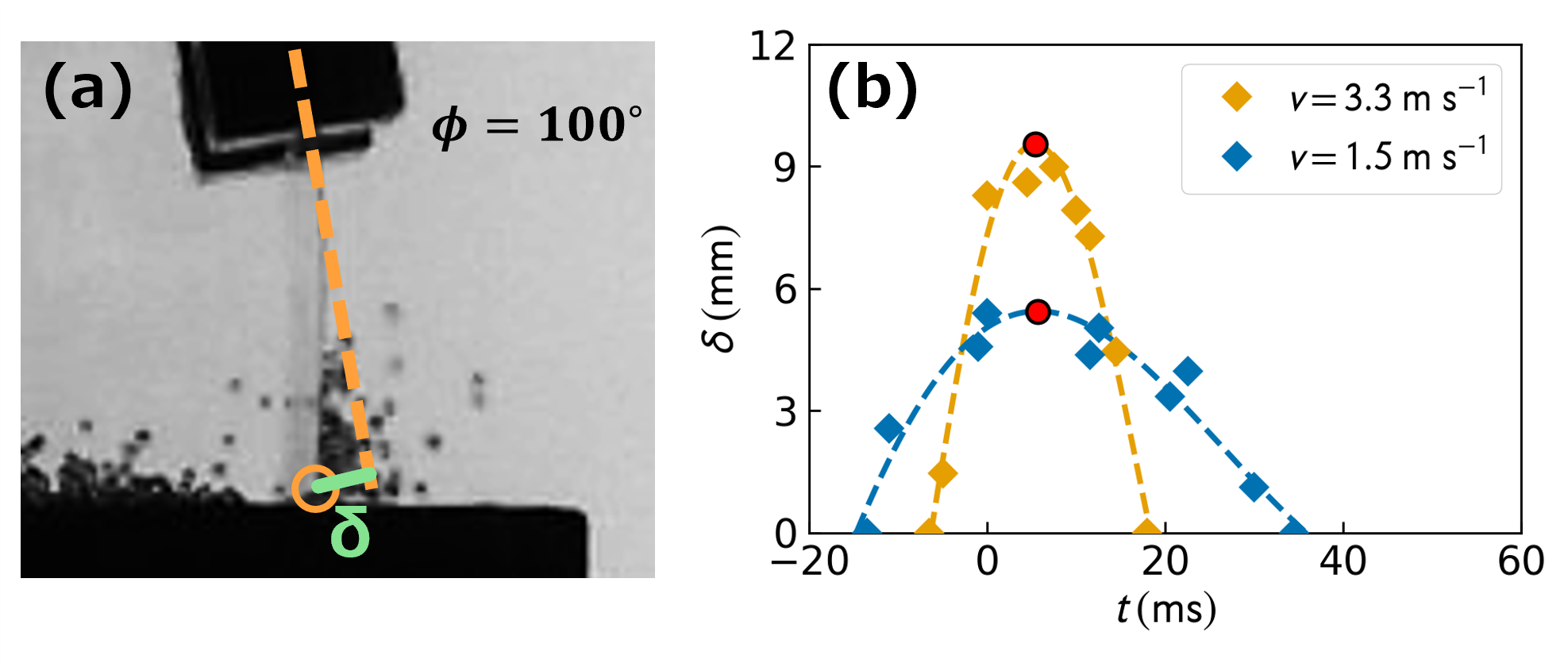}
    \caption{Plate deformation. (a)~Definition of $\delta$ as the perpendicular distance from the deflected plate tip to the virtual extension of the undeformed plate. (b)~Temporal evolution of $\delta$ for the Flexible plate at two sweeping velocities ($v = 3.3$ and $1.5$~m~s$^{-1}$). Dashed curves are damped-oscillation fits. Red filled circles indicate $\delta_{\mathrm{max}}$.}
    \label{fig:delta}
\end{figure}

\textbf{Definition of $\delta$.} The plate deformation $\delta$ is defined as the perpendicular distance from the deflected plate tip to the virtual extension of the undeformed plate, as illustrated in Fig.~\ref{fig:delta}(a). The plate tip was tracked frame by frame in the high-speed images. This definition separates the bending of the plate from the rigid-body motion of the pendulum arm.

\textbf{Extraction of $\delta_{\mathrm{max}}$.} Fig.~\ref{fig:delta}(b) shows the temporal evolution of $\delta$ for the Flexible plate at two sweeping velocities. The plate bends as it enters the granular bed, reaches a maximum deflection, and then relaxes with a damped oscillation after leaving the bed. Near the peak, $\delta(t)$ was fitted with a damped-oscillation function,
\[\delta(t) = \delta_0 \, e^{-t/\tau} \cos(\omega t + \varphi) + c,\]
around the peak time. Here, $\delta_0$, $\tau$, $\omega$, $\varphi$, and $c$ are fitting parameters. The maximum deformation $\delta_{\mathrm{max}}$ was taken as the peak value of the fitted curve, indicated by the red filled circles in Fig.~\ref{fig:delta}(b). The response consists of a single peak followed by a damped oscillation, supporting the mass-spring treatment of the plate used in the main text.

\section{Ejection speed \texorpdfstring{$V$}{V} measurement}
\label{sec:V}
\textbf{Image preprocessing and binarization.} For each captured frame, foreground signals representing the ejected particles were isolated by recursively removing background noise. Pixels falling within $2.5$ standard deviations ($2.5\sigma$) of the primary background intensity peak were filtered out based on global intensity histograms. Subsequently, the pixel intensity threshold for the ejecta was determined by identifying the most prominent peak of the remaining signal. The boundaries of the particle cloud were then defined as the points where the average pixel intensity dropped below $2\%$ of the peak intensity height.

\textbf{Dynamic region of interest (ROI).} The onset of the image analysis was defined as the frame immediately preceding the entry of the plate tip into a designated rectangular tracking zone, located near the point of particle-plate separation. The subsequent motion of the particle cloud was tracked for a duration of approximately $9.5$~ms post-separation. To reliably track the expanding cluster, we implemented a dynamic ROI whose left boundary shifts horizontally in synchronization with the pendulum's forward sweep. This prevents the trailing pendulum arm from re-entering the frame and triggering false detections.

\begin{figure}[htbp]
    \centering
    \includegraphics[width=1\linewidth]{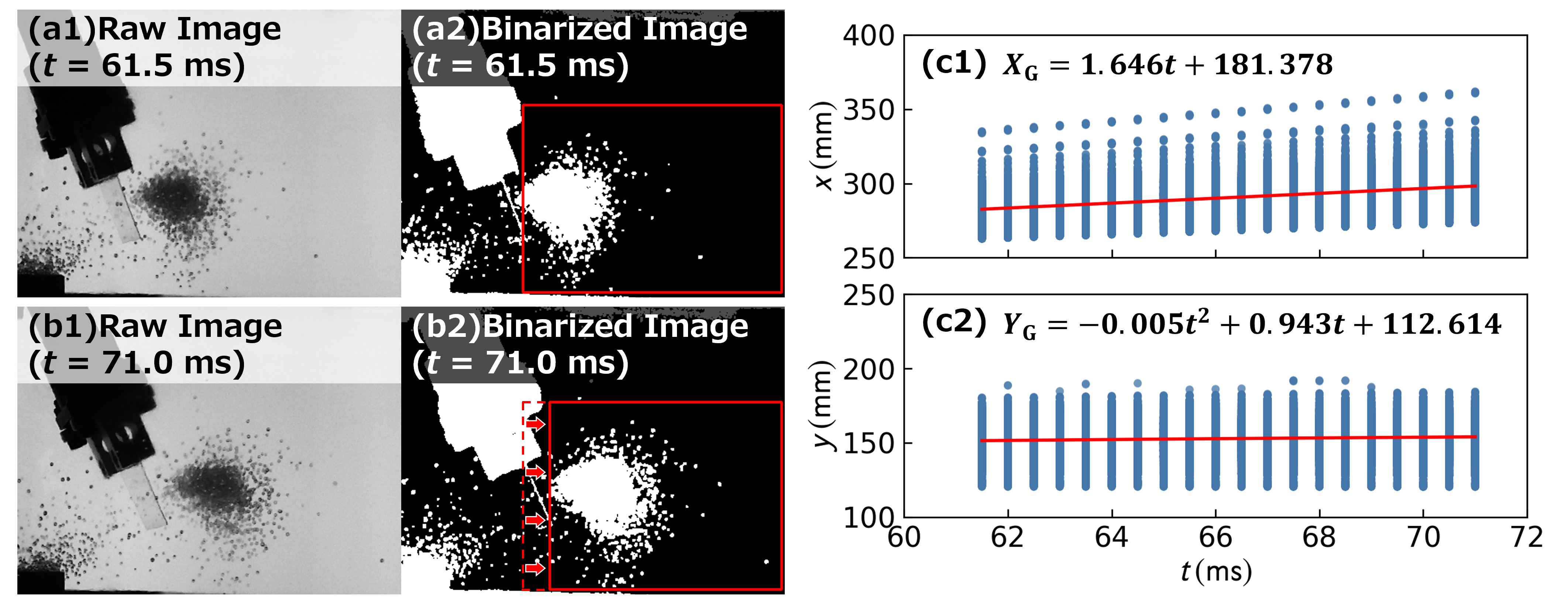}
    \caption{Image analysis procedure of particle ejection behavior. (a1, b1)~Raw high-speed images captured at $t = 61.5$~ms (analysis onset) and $71.0$~ms (termination) after $\phi = 90^\circ$ for the case of $D_{\mathrm{p}} = 1.0$~mm using the Flexible plate at a sweeping velocity $v \simeq 1.55$~m~s$^{-1}$. (a2, b2)~Binarized images corresponding to the raw frames, where the red rectangles indicate the analysis ROI. The left boundary of the ROI in (b2) is dynamically translated to block the intrusion of the moving plate. (c1, c2)~Spatiotemporal evolution of the ejected particle coordinates. Each blue point represents an individual bright pixel detected within the ROI, with its vertical spread capturing the spatial dilation of the particle cloud. A global least-squares fit applied to the entire point cloud in all frames yields the center-of-mass trajectory ($X_{\mathrm{G}}$, $Y_{\mathrm{G}}$).}
    \label{fig:v}
\end{figure}

\textbf{Point-cloud regression and velocity calculation.} The macroscopic ejection velocity $V$ was evaluated using a spatiotemporal point-cloud regression method that incorporates the coordinates of all detected bright pixels simultaneously. Assuming ideal projectile motion under gravity, a direct global least-squares fit was applied to the coordinate dataset. The horizontal positions $x_i(t)$ and vertical positions $y_i(t)$ of the $i$-th particle were modeled using linear and quadratic functions, respectively,
\[x_i(t) = V_x t + x_0,\]
\[y_i(t) = a t^2 + V_y t + y_0,\]
where $a = -g/2$ is fixed by the gravitational acceleration (expressed in mm~ms$^{-2}$). The collective initial velocity components $V_x$ and $V_y$ were extracted directly from these fitting coefficients. This global fit effectively captures the center-of-mass trajectory ($X_{\mathrm{G}}, Y_{\mathrm{G}}$) of the expanding particle cloud. Finally, the representative ejection speed was determined as $V = \sqrt{V_x^2 + V_y^2}$. The entire procedure is illustrated in Fig.~\ref{fig:v}.

\section{Estimation of the mobilized granular mass \texorpdfstring{$M_{\mathrm{add}}$}{Madd}}
\label{sec:Madd}
The elastic response time $t_{\mathrm{plate}}$ defined in the main text requires the granular mass that loads the plate. We write this mass in the general form,
\[M_{\mathrm{add}} \sim \rho w \ell d,\]
where $w$ is the plate width, $d$ is the penetration depth, and $\ell$ is the forward extent of the region set into motion. The three lengths are the transverse span of the contact, the forward reach of the disturbance, and the thickness of the mobilized layer.

Substituting this form into the definitions of $t_{\mathrm{plate}}$ and $t_{\mathrm{sweep}}$ in the main text gives
\[\mathrm{Sk} \sim \sqrt{\frac{\rho v^2 w \ell L^3}{EI R}}.\]
The penetration depth cancels for any $\ell$ that is itself independent of $d$. This separates two options.

If the forward reach is set by the sweep, the natural choices are $\ell \sim \sqrt{Rd}$ and $\ell \sim d$, and Sk retains a factor $d^{1/4}$ or $d^{1/2}$. If instead the forward reach is set by the plate, $\ell$ carries no $d$ and the depth cancels. The data of Fig.~\ref{fig:MV} of the main text do not show a significant difference between the two depths, which is consistent with the plate-set option. The result is reasonable as long as we consider the shallow sweeping limit.

\begin{figure}[htbp]
    \centering
    \includegraphics[width=0.3\linewidth]{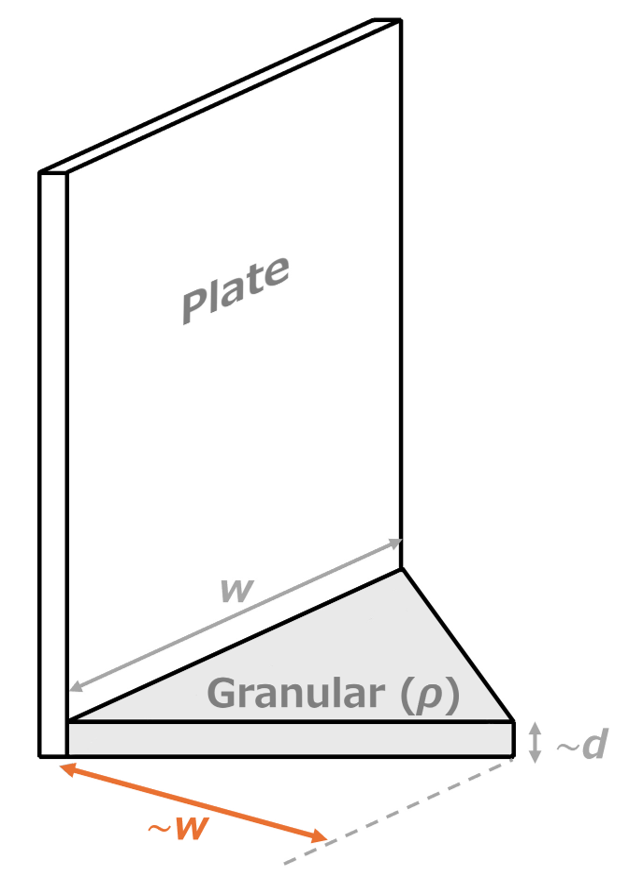}
    \caption{Schematic of the forward influence range. A plate of width $w$ mobilizes $M_{\mathrm{add}} \sim \rho w^2 d$ within a region of order $w$. The triangle shape has no particular meaning. Any shape with a forward extent of order $w$ gives the same scaling.}
    \label{fig:madd}
\end{figure}

Then, $\ell \sim w$ is naturally defined,
\[M_{\mathrm{add}} \sim \rho w^2 d,\]
as sketched in Fig.~\ref{fig:madd}. Only the scaling is retained. The geometry-dependent prefactor of order unity is absorbed into the definition of the skimming number. A simple plate geometry is used throughout, so the present data do not separate $\ell \sim w$ from $\ell \sim L$. Both give the same cancellation of $d$ and differ here only by a fixed factor. Physically, $\ell$ should not grow without bound as $L$ increases. Moreover, internal friction transmits stress laterally over a distance comparable to $w$. Thus, the current form is physically reasonable.

\section{Geometric reference mass \texorpdfstring{$M_{\mathrm{geo}}$}{Mgeo}}
\label{sec:Mgeo}
The plate tip traces an arc of radius $R$ while the granular surface remains flat, so the swept cross section is a circular segment of depth $d$. Its area is $R^2\cos^{-1}(1-d/R) - (R-d)\sqrt{2Rd-d^2}$. Multiplying this by the plate width $w$ and taking the shallow-sweeping limit ($d \ll R$) gives
\[
M_{\mathrm{geo}} = \rho w d \sqrt{Rd}.
\]
While $M_\mathrm{add}$ reflects the instantaneous mass effective for elastic oscillation, $M_{\mathrm{geo}}$ represents the accumulated scooped mass. $M_{\mathrm{geo}}$ counts only the material inside the swept segment. Grains lying ahead of that segment can also be driven forward and ejected, so $M/M_{\mathrm{geo}}$ is a normalized transport efficiency and is not bounded by unity.

\bibliography{GSI}

\end{document}